\documentclass[numbers,sort&compress]{IntechOpen-Book}
\usepackage{booktabs}

\graphicspath{{Artworks/}}

\begin{document}

\Mainmatter

\begin{frontmatter}

\chapter{Bidirectional Human–AI Alignment in Education for Trustworthy Learning Environments}
\author{Hua Shen, NYU Shanghai, New York University}

\makechaptertitle


\chaptermark{Bidirectional Human–AI Alignment in Education for Trustworthy Learning Environments}

\begin{abstract} 
Artificial intelligence (AI) is transforming education, offering unprecedented opportunities to personalize learning, enhance assessment, and support educators. Yet these opportunities also introduce risks related to equity, privacy, and student autonomy. This chapter develops the concept of bidirectional human–AI alignment in education, emphasizing that trustworthy learning environments arise not only from embedding human values into AI systems but also from equipping teachers, students, and institutions with the skills to interpret, critique, and guide these technologies. Drawing on emerging research and practical case examples, we explore AI’s evolution from support tool to collaborative partner, highlighting its impacts on teacher roles, student agency, and institutional governance. We propose actionable strategies for policymakers, developers, and educators to ensure that AI advances equity, transparency, and human flourishing rather than eroding them. By reframing AI adoption as an ongoing process of mutual adaptation, the chapter envisions a future in which humans and intelligent systems learn, innovate, and grow together.


\end{abstract}

\begin{keywords} 
AI for Education, Human-AI Alignment, Evolving AI Roles in Education
\end{keywords}

\end{frontmatter}

\section{Introduction} 
%


Artificial intelligence (AI) is rapidly becoming woven into the fabric of education~\cite{chap01:bib01,chap01:bib02}. From adaptive learning platforms and automated assessment tools to intelligent tutoring systems and predictive analytics, AI technologies promise to reshape how students learn, how teachers teach, and how schools make decisions~\cite{chap01:bib03,chap01:bib04}. These innovations hold the potential to address long-standing challenges—personalizing learning at scale, reducing administrative burdens, and generating timely insights to support both teaching and student progress.

Yet alongside this promise lie significant risks. Without deliberate attention to ethics, inclusivity, and transparency, AI systems may amplify biases, erode student privacy, or diminish teacher autonomy. They can also disrupt the delicate social dynamics of classrooms, altering power relationships between students, educators, and technology providers. In this sense, the introduction of AI into education is not merely a technical shift but a cultural and normative one~\cite{chap01:bib05}.

Building on the ``Bidirectional Human-AI Alignment'' framework~\cite{chap01:bib06,chap01:bib07}, this chapter proposes a bidirectional approach to human–AI alignment in education—one that treats alignment not as a one-way imposition of human values onto machines, but as a dynamic process in which both humans and AI systems co-adapt. On one hand, AI must be designed and deployed to reflect shared educational values such as equity, trust, and student agency~\cite{chap01:bib08,chap01:bib10}. On the other hand, educators and learners must also develop new literacies, skills, and mindsets to engage productively and critically with AI systems~\cite{chap01:bib09}. Only through this reciprocal relationship can AI enhance, rather than undermine, the human goals of learning.
The chapter unfolds in five major parts:


\begin{itemize}
    \item \textbf{What Needs to Be Aligned} – We begin by unpacking the foundations of alignment in education: shared values and ethical principles, clear learning goals, and well-defined interaction norms between humans and AI systems.

    \item \textbf{Pathways to Achieving Alignment} – We explore technical design strategies, ethical and legal frameworks, and mechanisms for continuous feedback and adaptation that sustain alignment over time.

    \item \textbf{Evolving Roles of AI in Education} – We trace AI’s trajectory from support tool to collaborative partner, focusing on adaptive learning, personalization, and enhanced assessment.

    \item \textbf{Impacts of AI on the Educational Ecosystem} – We examine how AI affects teacher roles, student agency, classroom safety, and the broader capacity to interpret and critique algorithmic decisions.

    \item \textbf{Moving Forward: Actions and Recommendations} – We conclude with practical steps for policymakers, educators, and developers, and identify key areas for future research and innovation.

\end{itemize}

By weaving these threads together, the chapter aims to provide a roadmap for designing trustworthy learning environments where human and artificial agents work in concert. Our goal is not simply to safeguard against harm, but to actively cultivate a culture of mutual trust, transparency, and shared responsibility—ensuring that AI enhances the human dimensions of education rather than eroding them.

\section{What Needs to Be Aligned?}
The promise of AI in education hinges on alignment—ensuring that technologies support, rather than distort, the aims of teaching and learning. Alignment is not a single dimension; it is a multi-layered process spanning values, goals, and norms~\cite{chap01:bib06}. Before technical solutions or policy frameworks can be implemented, we must first clarify \textbf{\emph{what}} should be aligned~\cite{chap02:bib03}. This section identifies three foundational elements: (1) core values and ethical principles, (2) educational goals and desired outcomes, and (3) human–AI interaction norms and boundaries.

\subsection{Core Values and Ethical Principles}

Trustworthy learning environments are grounded in shared values and ethical principles~\cite{chap02:bib04,chap02:bib05}. Drawing on human value theories from social science and psychology (e.g., Schwartz’s theory of basic values~\cite{chap02:bib01,chap02:bib02}), core principles such as equity, inclusivity, privacy, transparency, accountability, and respect for human dignity should serve as the “north star” for designing and deploying AI in education~\cite{chap02:bib06,chap02:bib07,chap02:bib08}. Without an explicit commitment to these principles, even well-intentioned technologies risk reinforcing biases, commodifying learning, or eroding student agency~\cite{chap02:bib09}.
I elaborate below on several key values that are particularly important for AI in education. 

\textbf{Equity and Inclusion}. AI systems should reduce, not exacerbate, disparities in access, achievement, and representation. This requires diverse training data, inclusive design processes, and ongoing monitoring for unintended discrimination.

\textbf{Privacy and Data Protection}. Given the sensitivity of student data, privacy safeguards and transparent data governance are non-negotiable.

\textbf{Transparency and Explainability}: Educators, students, and parents should be able to understand how AI systems reach their recommendations or decisions.

\textbf{Accountability}. Clear lines of responsibility must be established so that when systems err, human actors—not algorithms—are ultimately answerable.

Articulating and embedding these principles early in design can provide a moral compass for developers and institutions, ensuring technology supports the flourishing of learners rather than merely optimizing measurable outcomes.

\subsection{Educational Goals and Desired Outcomes}

Aligning AI with ethical values is only one part of the equation; it must also align with the substantive goals of education itself~\cite{chap02:bib10}. Educational systems are built not only to deliver content but to nurture critical thinking, creativity, collaboration, and lifelong learning. AI should therefore be evaluated not just on efficiency gains but on its contribution to these broader aims. 
Integrating AI into education requires careful consideration of its effects on curricula, pedagogy, and learner development. Key dimensions include:

\textbf{Curricular Alignment.} AI should reinforce intended learning objectives rather than introduce incentives that distort student focus. Systems that reward superficial engagement risk undermining deeper understanding and conceptual mastery.

\textbf{Skill Development.} Beyond academic knowledge, AI should support higher-order skills such as problem-solving, creativity, digital literacy, and ethical reasoning. Tools that simulate real-world challenges or collaborative projects can foster these transferable skills.

\textbf{Student Agency.} Effective education empowers learners to set goals, make choices, and reflect on progress. AI can facilitate this through adaptive pathways and meaningful feedback, while avoiding approaches that encourage passive consumption.

\textbf{Holistic Development.} Education extends to social-emotional learning, well-being, and civic responsibility. AI should support these dimensions—for example, by prompting reflection, fostering empathy, or promoting collaborative civic projects—rather than marginalizing them.

By making desired outcomes explicit, can better judge whether an AI product is a true pedagogical asset or merely a novelty.
%
True pedagogical alignment requires moving beyond efficiency gains to support multidimensional growth, ensuring AI serves as a partner in human development rather than a substitute for the essential human elements of education.




\subsection{Human–AI Interaction Norms and Boundaries}

Aligning AI with educational goals requires clear norms and boundaries for human–AI interaction. These define expectations for transparency, consent, roles, and the appropriate limits of automation, ensuring AI supports rather than undermines teaching and learning.

\textbf{Transparency of Roles.} Students and educators should clearly understand when they are interacting with AI, what data is collected, and how outputs are generated. Transparency fosters trust and informed engagement.

\textbf{Human Oversight.} Decisions with significant consequences—such as grading, placement, or disciplinary actions—must remain under human review to preserve fairness, contextual judgment, and accountability.

\textbf{Boundaries of Influence.} AI should augment, not replace, teacher judgment or peer interaction. Well-defined limits prevent over-automation that could erode essential human relationships in learning.

\textbf{Norms of Conduct.} Ethical and respectful behavior applies to both humans and AI systems. Tutors, chatbots, and other AI tools should model constructive, respectful communication.

\textbf{Reciprocal Learning.} As AI adapts to learners, students and teachers must also develop the literacy to critique, guide, and influence these systems, fostering a two-way learning process.

Establishing these norms and boundaries creates a shared framework of trust, clarifying which aspects of teaching and learning can be safely automated and which must remain inherently human.

Together, these three elements—core values, educational goals, and interaction norms—define the foundation for bidirectional human–AI alignment. Without clarity in these areas, even sophisticated technical or policy measures risk addressing symptoms rather than causes.

\section{Pathways to Achieving Alignment}
Clarifying what needs to be aligned is only the first step~\cite{chap03:bib04}. The next challenge is to translate those principles, goals, and norms into practice~\cite{chap03:bib03}. This section outlines three complementary pathways—technical approaches and design strategies, ethical/legal/policy frameworks, and ongoing evaluation with feedback and adaptation—that together create an ecosystem for trustworthy AI in education.

\subsection{Technical Approaches and Design Strategies}

Embedding alignment into AI systems begins at the design stage~\cite{chap03:bib05}. Rather than treating ethics or usability as add-ons, they must be integral to the development lifecycle.

\begin{itemize}
    \item \textbf{Human-Centered Design}: Engage educators, students, and parents early and throughout the design process to ensure the system reflects real-world needs and values.
    \item \textbf{Value-Sensitive Design}: Translate ethical principles—fairness, privacy, transparency—into explicit technical requirements. For example, using explainable models where interpretability is critical.
    \item \textbf{Bias Mitigation and Inclusive Data}: Curate diverse datasets, run fairness audits, and employ algorithmic techniques to detect and reduce biases that could harm marginalized groups.
    \item \textbf{Personalization with Guardrails}: While adaptive learning systems tailor content to individual students, they should also respect boundaries, avoid over-surveillance, and maintain curricular integrity.
    \item \textbf{Transparency Features}: Dashboards, explanations, and open documentation can make it easier for teachers and students to understand why a system recommends a particular action.

\end{itemize}

Technical strategies alone cannot guarantee trustworthy systems, but they provide the “hardwiring” that makes alignment feasible at scale.

\subsection{Ethical, Legal, and Policy Frameworks}

Technical fixes must be supported by institutional and societal safeguards~\cite{chap03:bib06,chap03:bib07,chap03:bib08}. Ethical, legal, and policy frameworks provide the rules of the road that guide both developers and users~\cite{chap03:bib01,chap03:bib02}. Corresponding the core alignment objectives mentioned above, I discuss potential frameworks that are valuable for embedding values into AI systems.

\emph{Codes of Practice and Standards} indicate that professional associations and education ministries can publish clear standards on responsible AI use, data governance, and transparency.

\emph{Privacy and Data Protection Laws} emphasize that strong regulations around data collection, consent, and retention (e.g., GDPR-like policies) help ensure student data is handled ethically.

\emph{Procurement and Certification} show that schools and governments can require third-party audits or certifications of AI products before adoption, much like safety checks in other industries.

\emph{Accountability Mechanisms} require policies to define who is responsible when AI systems err—vendors, schools, or both—and establish processes for redress.

\emph{Equity Mandates} enable policy frameworks that can explicitly require equity impact assessments to prevent exacerbating digital divides.

These frameworks not only protect stakeholders but also create predictable conditions for innovation, allowing developers to build with confidence and schools to adopt with trust.

    
    


\subsection{Ongoing Evaluation, Feedback, and Adaptation}

Alignment is not a one-time achievement but a continuous process~\cite{chap03:bib09,chap03:bib10}. As classrooms evolve and AI systems learn, the conditions under which they operate also change. Continuous evaluation ensures systems remain fit for purpose.

\textbf{Monitoring and Auditing.} Regular audits of performance, fairness, and privacy compliance can catch issues before they escalate.

\textbf{Feedback Loops.} Provide clear channels for teachers, students, and parents to report problems or suggest improvements, and ensure those inputs feed back into system updates.

\textbf{Adaptive Governance.} Policies and norms should be revisited periodically to reflect new research, technologies, and societal expectations.

\textbf{Impact Studies.} Longitudinal research on AI’s effects on learning outcomes, student agency, and teacher roles can inform future design and policy.

\textbf{Professional Development.} Equip educators with ongoing training to understand updates, interpret outputs, and integrate AI tools responsibly.

This iterative approach mirrors how learning itself works—testing, feedback, and revision—ensuring that alignment remains robust over time.

Together, these pathways create a resilient infrastructure for bidirectional human–AI alignment in education. Technical design, governance frameworks, and continuous evaluation form a mutually reinforcing system: each supports the others to ensure AI remains trustworthy, and responsible to human needs.




\section{Evolving Roles of AI in Education}
AI in education has moved beyond experimental pilots to widespread adoption in classrooms, administrative systems, and educational platforms~\cite{chap04:bib01,chap04:bib02}. As these technologies mature, their roles evolve from simple tools to complex partners in the learning process~\cite{chap04:bib03,chap04:bib04}. This section traces that trajectory, highlighting how AI can (1) transition from support tool to collaborative partner, (2) enable adaptive and personalized learning, and (3) enhance assessment, feedback, and learning analytics.

\subsection{From Support Tool to Collaborative Partner to Interactive Tutor}

AI in education has evolved from performing narrowly defined, transactional tasks—such as grading multiple-choice tests, scheduling classes, or delivering drill-and-practice exercises—toward becoming a more relational and collaborative presence in learning environments~\cite{chap04:bib05}. This evolution reflects a continuum in AI’s role, from support tool to collaborative partner to interactive tutor (Table~\ref{tab:ai_roles_compact}).

\begin{table}[h!]
\centering
\footnotesize
\caption{Roles of AI in Education: Functions, Features, and Implications}
\label{tab:ai_roles_compact}
\begin{tabular}{@{}p{1.5cm}p{2.5cm}p{3.5cm}p{4cm}@{}}
\toprule
\textbf{Role} & \textbf{Primary Function} & \textbf{Typical Features} & \textbf{Implications for Educators \& Students} \\ 
\midrule
\textbf{Support Tool} & Automates routine or administrative tasks & Grading multiple-choice tests, scheduling, drill-and-practice exercises & Frees educators to focus on higher-level instruction; limited relational interaction with students \\

\textbf{Collaborative Partner} & Assists teachers in decision-making and planning & Lesson design, identifying at-risk students, differentiated instruction, teacher dashboards, analytics & Enables co-piloting with AI; supports teacher collaboration and data-driven interventions; requires trust and understanding of AI recommendations \\

\textbf{AI Tutor} & Provides adaptive, personalized learning support directly to students & One-on-one dialogue, adaptive hints, tailored explanations, co-creation of exercises and simulations & Offers individualized guidance; enhances student engagement and learning; requires clear boundaries and oversight to preserve human judgment \\
\bottomrule
\end{tabular}
\end{table}

\textbf{Support Tool.} In its initial form, AI primarily handled administrative or repetitive tasks, providing efficiency and freeing educators to focus on higher-level instructional work.

\textbf{Collaborative Partner.} AI increasingly engages educators in shared decision-making. Systems can assist in designing lesson plans, identifying at-risk students, and differentiating instruction, transforming teachers into co-pilots rather than passive users. AI-powered dashboards and analytics further facilitate teacher collaboration, enabling data-driven interventions and coordinated instructional planning.

\textbf{Interactive Tutor.} Advanced AI systems now simulate one-on-one tutoring, engaging students in adaptive dialogue, asking probing questions, offering hints, and tailoring explanations to individual learning responses. Some platforms even support co-creation of learning content, generating exercises, simulations, or multimodal experiences that blend human creativity with computational power.

This progression—from tool to partner to tutor—raises important considerations regarding trust, transparency, and autonomy. As AI assumes more relational and pedagogical roles, educators and students must establish clear expectations about responsibilities, boundaries, and oversight to ensure AI enhances learning without undermining essential human judgment.




\subsection{Enabling Adaptive and Personalized Learning}

A central promise of AI in education lies in its ability to deliver personalized learning at scale—an aspiration that remains challenging for even the most skilled educators managing large, diverse classrooms. AI-driven personalization can enhance engagement, accelerate learning, and support students’ individual needs through several key mechanisms:

\begin{itemize}
    \item \textbf{Dynamic Content Delivery}: AI systems can adjust lesson difficulty, sequencing, and modality (e.g., text, video, or simulation) to align with a learner’s current knowledge, skills, and preferences.
    \item \textbf{Real-Time Feedback}: Adaptive platforms provide immediate feedback, hints, or scaffolding, helping learners remain within their “zone of proximal development” and optimize growth.
    \item \textbf{Individualized Learning Pathways}: By recommending enrichment opportunities or targeted remediation, AI ensures that advanced learners are challenged while those requiring support receive tailored interventions.
    \item \textbf{Multimodal Accessibility}: Personalization extends beyond content pacing, incorporating language support, assistive technologies, and culturally relevant materials to promote inclusivity and equity.
\end{itemize}

While AI can substantially enhance personalization, its implementation must be carefully balanced with considerations of equity and learner autonomy. Systems that are overly prescriptive, opaque, or deterministic risk constraining students to fixed learning trajectories, potentially limiting exploration and self-directed growth. Educators play a critical role in maintaining oversight, fostering metacognitive awareness, and helping students understand the rationale behind personalized recommendations, thereby preserving agency and promoting deeper learning.




\subsection{Enhancing Assessment, Feedback, and Learning Analytics}




Assessment and feedback are central to the learning process, yet traditional approaches are often slow, resource-intensive, and limited in scope. AI has the potential to transform this landscape by providing more timely, nuanced, and actionable insights. For instance, AI systems can extend beyond grading multiple-choice or essay items to evaluate complex learning activities such as problem-solving processes, collaborative discussions, and oral presentations.

AI also enables continuous, low-stakes feedback through embedded micro-assessments. These assessments provide students with ongoing guidance throughout the learning process, reducing the pressure associated with high-stakes examinations while supporting incremental improvement. By aggregating data across learning activities, AI-driven learning analytics can reveal patterns and trends at the classroom, school, or district level, thereby informing targeted interventions and guiding the strategic allocation of educational resources.

Predictive analytics further enhance instructional decision-making by identifying students who may require additional support before traditional performance indicators signal concern. Coupled with well-designed dashboards and visualization tools, these analytics empower teachers, students, and parents to engage with data in a meaningful way, closing the feedback loop and making insights actionable rather than abstract.

Despite these benefits, the use of AI in assessment and analytics raises important concerns around privacy, consent, and interpretability. Data must always be contextualized, and analytics-driven recommendations should complement, rather than replace, the professional judgment and nuanced understanding that educators bring to their students’ learning needs.

Together, these evolving roles highlight AI’s dual nature in education—as a transformative enabler and a potential disruptor. By treating AI as a partner rather than a substitute, and by embedding transparency and human oversight, schools can harness these new capabilities to enhance rather than diminish the human experience of learning.

\section{Impacts of AI on the Educational Ecosystem}
As AI tools become more embedded in classrooms and educational institutions~\cite{chap05:bib01,chap05:bib02}, their influence extends beyond instructional delivery to the very structure of teaching, learning, and governance~\cite{chap05:bib03,chap05:bib04}. These changes affect not only what students learn but how educators work, how decisions are made, and how agency is distributed~\cite{chap05:bib05}. This section explores four critical dimensions of AI’s impact: (1) the human capacity to interpret and critique AI decisions, (2) AI’s role in promoting safety and well-being, (3) shifts in pedagogical practices and teacher roles, and (4) effects on student autonomy, agency, and motivation.

\subsection{Human Capacity to Interpret and Critique AI Decisions}

For AI to be trustworthy in education, teachers, students, and administrators must be able to interpret, question, and critique its outputs. Without this capacity, AI risks becoming a “black box” authority.

\begin{itemize}
    \item \textbf{Algorithmic Literacy}: Educators need professional development to understand how algorithms work, where biases may arise, and how to contextualize recommendations.
    \item \textbf{Transparency Tools}: Clear explanations, confidence scores, and data visualizations can help stakeholders see why an AI system produced a particular output.
    \item \textbf{Critical Pedagogy}: Students can be taught to question automated feedback and recommendations, integrating digital and algorithmic literacy into curricula.
    \item \textbf{Shared Decision-Making}: Systems should be designed so that human users can override or appeal AI recommendations, reinforcing human judgment as the final authority.
\end{itemize}

Building this interpretive capacity is essential for maintaining accountability, avoiding blind trust, and fostering a culture of informed skepticism rather than passive compliance.





\subsection{Shifts in Pedagogical Practices and Teacher Roles}

AI in education does more than automate routine tasks; it fundamentally reshapes pedagogical practices and the role of teachers. As AI systems increasingly deliver personalized content, the traditional focus on content transmission shifts toward facilitation. Teachers can devote more time to mentoring, project-based learning, and supporting students’ social-emotional development, emphasizing the human dimensions of education that AI cannot replicate.

Real-time analytics provide educators with actionable insights, enabling data-informed instruction. Teachers can adjust lesson pacing, groupings, and targeted interventions responsively, tailoring learning experiences to meet diverse student needs. This shift necessitates the development of new professional competencies, including data literacy, ethical oversight, and technological fluency, ensuring that educators are prepared to navigate AI-enhanced classrooms effectively.

In this evolving landscape, teachers increasingly function as orchestrators who collaborate with AI tools—much like pilots supervising an autopilot system. AI can provide scaffolding, suggestions, and monitoring, while teachers make nuanced judgments and maintain human oversight. However, without careful implementation, there is a risk of de-skilling, as reliance on AI for core instructional tasks may erode professional expertise. Schools must therefore ensure that AI augments, rather than replaces, teacher practice.

When integrated thoughtfully, these shifts empower educators to focus on the uniquely human aspects of teaching—empathy, mentorship, and creativity—while maintaining autonomy and professional judgment. Effective training and institutional support are essential to realizing the full potential of AI in reshaping pedagogy.



These shifts can empower teachers to focus on the uniquely human aspects of education—empathy, mentorship, and creativity—if they are supported with training and autonomy.

\subsection{Effects on Student Autonomy, Agency, and Motivation}

One of the most profound impacts of AI in education is its influence on students’ sense of control, motivation, and identity as learners.

\textbf{Enhanced Agency Through Personalization}. Adaptive systems can let students progress at their own pace, choose learning modalities, and receive tailored feedback—fostering ownership.

\textbf{Risks of Over-Scaffolding.} 
At the same time, there are risks associated with over-scaffolding. If AI systems become overly directive, they may reduce opportunities for struggle, experimentation, and self-regulation—skills that are critical for lifelong learning.

\textbf{Motivational Dynamics.} Similarly, gamified or algorithmically optimized environments can increase engagement, but they may also create reliance on extrinsic rewards, potentially undermining intrinsic motivation.

\textbf{Transparency and Choice.} To mitigate these risks, students should understand the rationale behind AI-generated content and recommendations and be empowered to adjust or opt out of suggested learning paths.

\textbf{Developing Metacognition.} 
Educators play a crucial role in leveraging AI feedback to cultivate metacognitive skills, guiding students to reflect on their learning strategies, build self-awareness, and develop resilience.

The challenge lies in striking a balance where AI empowers rather than constrains, amplifying students’ intrinsic motivation and supporting self-directed learning.

Together with changes in pedagogy and assessment, these developments illustrate that AI’s impact on education is both systemic and deeply personal. By anticipating these shifts and shaping their implementation proactively, policymakers, schools, and developers can ensure that AI enhances—rather than diminishes—the human capacities that are central to meaningful learning experiences.





\section{Moving Forward: Actions and Recommendations}
The preceding sections have demonstrated that AI in education carries both transformative potential and significant risks. Realizing its benefits while minimizing harms requires deliberate, coordinated action by policymakers, educators, institutions, developers, and researchers. This chapter concludes with practical recommendations under three broad areas: (1) policy guidelines for responsible AI integration, (2) best practices for educators, institutions, and developers, and (3) key priorities for future research and innovation.

\subsection{Policy Guidelines for Responsible AI Integration}

Robust policy frameworks can establish a foundation of trust and accountability that enables innovation while safeguarding learners. Transparency and accountability should be central: educational authorities and technology providers need to disclose the purposes, data sources, and performance metrics of AI systems in ways that are accessible to the public. Stronger privacy and data protection laws are essential to safeguard student information, including strict limits on data retention, sharing, and commercial exploitation.

Ethical and equity impact assessments should become standard practice before large-scale deployments, ensuring that AI systems do not inadvertently exacerbate disparities. Governments and accrediting bodies can also create certification schemes for AI education products—analogous to safety or accessibility ratings—so that schools can make informed purchasing and implementation decisions.

Interoperability and open standards are equally important. By promoting platforms that enable data portability and integration across multiple tools, policymakers can reduce vendor lock-in and allow for holistic oversight of students’ learning data. Finally, stakeholder engagement must be embedded into policy design. Regular consultation with teachers, parents, and students can ensure that regulations remain responsive to real-world needs and inclusive of diverse perspectives, rather than imposed from the top down.



\subsection{Best Practices for Educators, Institutions, and Developers}

While policy frameworks provide the foundation for responsible AI use, day-to-day practices within schools and development teams determine how these technologies actually affect teaching and learning. Educators, institutions, developers, and students each play a crucial role in ensuring AI tools enhance rather than diminish educational quality.

For \textbf{educators}, the priority is to integrate AI thoughtfully into pedagogical practice. Teachers should treat AI as a complement to—not a replacement for—human expertise. This means maintaining active oversight of AI recommendations, fostering students’ metacognitive skills, and using data insights to inform rather than dictate instruction. Professional development is essential: training in data literacy, ethical oversight, and technological fluency can help teachers navigate AI-enhanced classrooms with confidence.

\textbf{Institutions} can support responsible adoption by creating clear governance structures and ethical guidelines for AI use. School leaders should ensure transparency with parents and students about how AI systems work, what data they collect, and how that data will be used. Institutions can also invest in digital infrastructure that prioritizes security, interoperability, and accessibility, thereby enabling educators to experiment with AI tools without compromising privacy or equity.

\textbf{Developers}, meanwhile, should design AI systems with the educational context in mind. This includes building transparent interfaces, offering meaningful user control, and testing for bias or unintended consequences before deployment. Collaboration with educators and students during the design phase can produce tools that align with real classroom needs rather than imposing rigid technological solutions.

For \textbf{students}, responsible use of AI involves cultivating digital literacy, self-regulation, and a critical awareness of how AI systems shape their learning. Learners should be encouraged to view AI as a tool to support—not substitute—their own thinking, and to question or adjust recommendations when appropriate. Educators and institutions can foster this agency by explicitly teaching students how to interpret feedback, manage data privacy, and reflect on their learning strategies. When students understand how AI systems operate and develop the confidence to use them judiciously, they are better equipped to take ownership of their education and to transfer these skills to future workplaces and civic life.

Taken together, these practices create an ecosystem in which AI strengthens teaching and learning. When educators retain agency, institutions provide safeguards, developers prioritize ethical design, and students develop the skills to use AI responsibly, the technology can serve as a catalyst for innovation while upholding the values of equity, inclusion, and human development at the heart of education.















\subsection{Key Areas for Future Research and Innovation}

Beyond immediate policy and practice, the next decade of AI in education will be shaped by targeted research and sustained innovation. Several areas merit particular attention if AI is to enhance—not diminish—human potential in learning environments.

First, researchers should deepen our understanding of how AI affects learning outcomes, equity, and student well-being over time. Longitudinal studies can illuminate not only academic gains but also impacts on motivation, agency, and social-emotional development. This evidence base is critical for moving beyond anecdotal claims and toward empirically grounded policy decisions.

Second, there is a need for more transparent and explainable AI systems tailored to educational contexts. Innovations in interpretable models and user-friendly dashboards can help educators, students, and parents understand why certain recommendations are made, strengthening trust and enabling informed decision-making.

Third, future development should prioritize inclusivity and cultural responsiveness. This involves building systems that work across languages, learning differences, and socio-economic contexts, while actively testing for and mitigating bias. Collaborative research between technologists, educators, and communities can ensure that tools are designed for diverse learners rather than assuming a single “typical” user.

Finally, the field would benefit from new paradigms that blend human and artificial intelligence in genuinely complementary ways. This includes co-creative learning environments, adaptive assessment models that foster metacognition, and hybrid teaching roles that emphasize empathy, mentorship, and creativity while drawing on AI’s data-processing strengths.

By investing in these areas, policymakers, institutions, and developers can help steer AI in education toward a future that prioritizes human flourishing. Research and innovation should not simply chase technical possibility but actively shape systems that embody educational values, safeguard equity, and cultivate lifelong learning.

\section{Conclusion}
This chapter has examined the foundations, pathways, and impacts of bidirectional human–AI alignment in education. We began by identifying what must be aligned—core values and ethical principles, educational goals, and human–AI interaction norms. We then mapped the pathways to achieve alignment through technical design strategies, ethical and legal frameworks, and ongoing evaluation. We traced AI’s evolving role from support tool to collaborative partner, explored how personalization and analytics can transform learning, and reflected on the implications for teacher practice, student agency, and institutional systems.

The central insight is that alignment is not a one-time act but an ongoing process. It requires translating principles into design, embedding safeguards into policy, and cultivating human capacity to interpret, critique, and improve AI systems. When done well, alignment enables AI to expand human potential, strengthen trust, and advance education’s deepest aims. When neglected, it risks undermining privacy, equity, and autonomy.

Looking ahead, the goal is not simply to “fit AI into schools” but to reimagine schools as spaces where humans and intelligent systems learn, adapt, and improve together. This vision frames AI not as a substitute for educators but as a collaborator that augments teaching, supports student growth, and creates new opportunities for creativity and critical thinking. Such synergy depends on mutual adaptation: AI systems designed for transparency, fairness, and inclusivity, and human actors equipped with the literacy and agency to guide, challenge, and improve those systems.

Achieving this vision demands collective action across all stakeholders. Policymakers must create clear guardrails and incentives for responsible innovation. Educators and institutions must develop new competencies, governance structures, and professional cultures. Developers must embed human values into the technical core of their products. Researchers must produce evidence and new models to anticipate risks and inform practice. And students themselves must be engaged as active co-designers and evaluators of the tools shaping their learning experiences.

By working together across these boundaries, we can create trustworthy learning environments where technology enhances—rather than diminishes—the human experience of education. This is the promise, and the responsibility, of bidirectional human–AI alignment: to ensure that as AI grows more powerful, education grows more equitable, empowering, and humane.

\end{document}